\documentstyle[twoside,fleqn,espcrc2,epsf]{article}
\title{Spin, Charge, and Bonding in Transition Metal Mono Silicides}
\author{D. van der Marel, A. Damascelli, K. Schulte
        \address{Solid State Physics Laboratory,
                 University of Groningen,
                 Groningen 9747 AG, The Netherlands}
         A. A. Menovsky
        \address{University of Amsterdam, 1018 XE Amsterdam, The Netherlands}}
\date{\today}

\begin{document}

\begin{abstract}\noindent
We review some of the relevant physical properties of 
the transition metal mono-silicides with the FeSi structure 
(CrSi, MnSi, FeSi, CoSi, NiSi, etc) and explore the relation between their 
structural characteristics and the electronic properties. We confirm the
suggestion orginally made by Pauling, that
the FeSi structure supports two quasi-atomic d-states at the 
transition metal atom. This shell contains from 0 to 4 
electrons in the sequence CrSi to NiSi.
In FeSi the two quasi-atomic d-electrons are responsible for
the high temperature $S=1$ state, which is compensated for T=0 by
two itinerant electrons associated with the Fe-Si resonance bonds.
\end{abstract}
\maketitle

\begin{table*}
\begin{displaymath}
\begin{array}{|l|l|l|l|l|l|l|l|l|}
\hline
              &a     &x_{TM}&1-x_{Si}&d_1 &d_2   &d_3      &d    &d_{sum}\\
              &(\AA) & (a)  & (a)   &(\AA) &(\AA) &(\AA) &(\AA) &(\AA) \\
\hline
 \mbox{ReSi}  &4.775 &0.14  &0.16  & 2.5   & 2.5   & 2.7   &2.56 & 2.55   \\
 \mbox{OsSi}  &4.728 &0.128 &0.161 & 2.364 & 2.429 & 2.715 &2.54 & 2.52   \\
 \mbox{RuSi}  &4.701 &0.128 &0.164 & 2.383 & 2.395 & 2.706 &2.53 & 2.51   \\
 \mbox{RhSi}  &4.676 &0.147 &0.155 & 2.451 & 2.480 & 2.548 &2.51 & 2.52   \\
 \mbox{CrSi}  &4.629 &0.136 &0.154 & 2.325 & 2.435 & 2.589 &2.49 & 2.45   \\
 \mbox{MnSi}  &4.558 &0.138 &0.154 & 2.305 & 2.402 & 2.537 &2.45 & 2.43   \\
 \mbox{FeSi}  &4.493 &0.136 &0.156 & 2.271 & 2.352 & 2.519 &2.41 & 2.44   \\
 \mbox{CoSi}  &4.447 &0.14? &0.16? & 2.2?  & 2.3?  & 2.5?  &2.4? & 2.42   \\
 \mbox{NiSi}  &4.437 &0.14? &0.16? & 2.2?  & 2.3?  & 2.5?  &2.4? & 2.42   \\
\hline
\end{array}
\end{displaymath}
\caption{Lattice parameter (a), 
displacement vectors x (units of lattice constant a), 
interatomic distances between TM and Si atoms ($d_1,d_2,d_3$), weighted 
average of the latter ($d$) and the sum of the atomic radii for 
12-coordinated metals and the single bond radius of Si ($d_{sum}$). 
For CoSi and NiSi no published values of x could be located.}
\label{table:radii}
\end{table*}
\section{Introduction}
Among the large variety of unusual properties found in transition metal
compounds, the phenomenon of spin-singlet formation is one of
the most intrigueing, and perhaps the most widely spread. The impact of
singlet formation on the physical state ranges
from insulating behaviour to superconductivity. In some cases
singlet formation is accompanied by a spin-Peierls lattice deformation, 
such as in CuGeO$_3$. In other cases the formation of singlets is accompanied
by a transition from a metallic to an insulating state. It has been suggested 
that FeSi is such a Kondo-insulator\cite{aeppli,mandrus,park}. In the present paper we discuss the wider 
class of transition metal mono-silicides (CrSi, MnSi, FeSi, CoSi, NiSi, etc)
with the FeSi structure, we review the structural and physical data of 
this class of compounds, and explore the relation between their structural
characteristics and the electronic properties.
\begin{figure}
\begin{center}
\leavevmode
\vspace{1cm}
\hbox{%
\epsfxsize=75mm
\epsffile{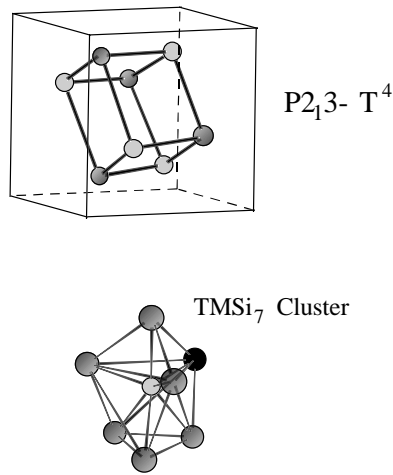}}
\end{center}
\vspace{0mm}
\caption{Crystal structure with the $T^4-P2_13$ space group, and a TMSi$_{7}$ cluster representing
the site symmetry of individual atoms in this structure.}
\label{fig:lattice}
\end{figure}
\section{Structural properties}
In Table \ref{table:radii} we summarize the lattice structures of the mono-silicides 
for the various d-transition elements\cite{xtal}. For the 3d series the
$T^4-P2_13$ space group has been observed from CrSi to NiSi. For NiSi in 
addition the NiAs structure has been reported. The $T^4-P2_13$ space group 
has also been reported for RuSi, RhSi, ReSi and OsSi. 
TiSi, VSi, PdSi, IrSi and PtSi all have the MnP structure. 
\\
In this paper we restrict the discussion to the mono-silicides with the
$T^4-P2_13$ space group (displayed in Fig. \ref{fig:lattice}). This simple
cubic structure contains four TM atoms in the positions 
($x,x,x; x+\frac{1}{2}, \frac{1}{2}-x,1-x$; etcycl.) and four Si
atoms in an equivalent set.
\\
In Table \ref{table:radii} we summarize the cell-parameters and the 
interatomic distances\cite{xtal}. 
The four transition metal and silicon atoms occupy equivalent sites.
The site symmetry has only a three-fold
rotation axis, which we indicate throughout this paper as the local
z-axis. The TM-atom has a 7-fold coordination by Si atoms and vice versa: A single bond
pointing toward the Si-atom along the z-axis (type 1), and two groups of three 
indentical bonds (types 2 and 3) with azimuthal angles relative to the z-axis
of 0 (type 1) 74 (type 2) and 141 degrees (type 3).  
The two groups of type 2 and 3 are rotated around the local z-axis
with an angle of $24\pm 2$ degree relative to each other.
The four equivalent sites within the unit cell have 
the same sense of rotation, hence two isomeric structures (lefthanded
and righthanded) exist. As a result
phase/anti-phase boundaries of the two isomers can occur (also in
single crystals) which may will considerable influence on the transport
properties. Single domain crystals should exhibit optical
activity. These aspects of the transition metal monosilicides still await
further scrutiny. 
\\
\begin{figure}
\begin{center}
\leavevmode
\vspace{0cm}
\hbox{%
\epsfxsize=75mm
\epsffile{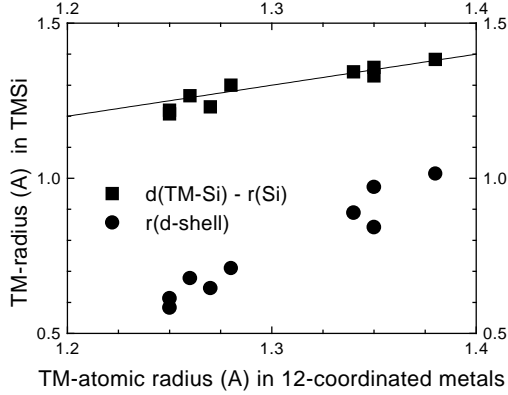}}
\end{center}
\vspace{0cm}
\caption{TM radius in TMSi (TM-Si bondlength minus radius of Si, squares) 
and d-shell radii (circles) versus the TM radius in 12-coordinated metals.}
\label{fig:radii}
\end{figure}
Within an accuracy of 2 \% the average distance between the TM-atom and its 7 
Si neighbours ($d$) follows the rule $d=r_{TM}+r_{Si}$,
where $r_{TM}$ is the atomic radius of the TM atom in 12-coordinated
metals\cite{pearson}, and $r_{Si}=1.173\AA$ is the single bond radius of Si 
(Fig. \ref{fig:radii} and Table \ref{table:radii}), 
indicating that almost no
charge transfer takes place between Si and the TM atom.
The absence of charge transfer between TM and Si can be anticipated from
the fact that the Pauling electronegativities are almost the same.
\\
For $x(TM) = 0.1545$ and $x(Si)= -0.1545$ all seven
neighbours have the same bondlength. As we can see in Table \ref{table:radii}
the structure of the TM-monosilicides is rather close to this situation,
but the deviations are significant. The variation of the
bondlength within the group of 7 nearest neighbours has been discussed
by Pauling\cite{pauling} for CrSi to NiSi in terms of valence bond sums, 
which lead to the following bond numbers:
one full bond (type 1), three 2/3 bonds(type 2) and three 1/3 bonds (type 3). 
\section{Electronic properties}
\subsection{Magnetic properties}
The best studied materials in this class are MnSi and FeSi. MnSi has
a helical spin structure at low fields below the Ne\`el temperature of
29 K, and is paramagnetic above $T_N$. CrSi and NiSi are either 
weakly paramagnetic or diamagnetic. Estimates of the local moment on Mn 
based on magnetic susceptibilities vary from 
$\mu_{eff}=2.12\mu_B$ with $g=2$ below 300 K to 3.3$\mu_B$ above
600 K\cite{foexradovskiisidorenko}. For FeSi\cite{jaccarino} the 
high temperature limiting behaviour of the susceptibility gives 
$\mu_{eff}=1.7$ with $g=3.9$, but good fits
to thermally activated Curie-Weiss behaviour were also obtained with
$\mu_{eff}=2.8$ and $g=2$. 
\subsection{Optical properties}
\begin{figure}
\vspace{0cm}
\hspace{0cm}
\begin{center}
\leavevmode
\hbox{%
\epsfxsize=80mm
\epsffile{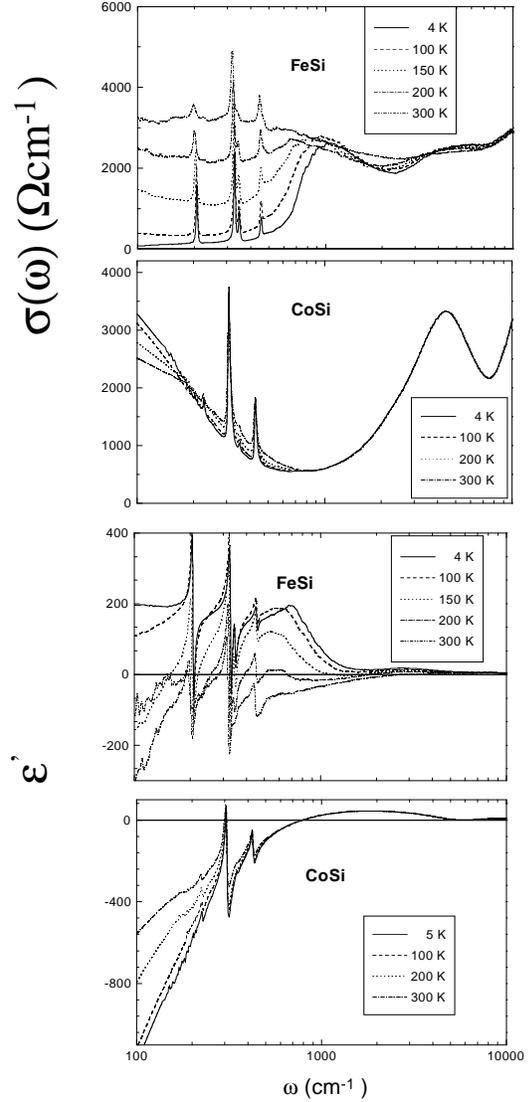}}
\end{center}
\vspace{0cm}
\caption{Real part of the optical conductivity ($\sigma_1$) and of the
dielectric constant ($\epsilon'$) of FeSi and CoSi.}
\label{fig:epsig.fesi.cosi}
\end{figure}
\begin{figure}
\begin{center}
\leavevmode
\vspace{0mm}
\hbox{%
\epsfxsize=75mm
\epsffile{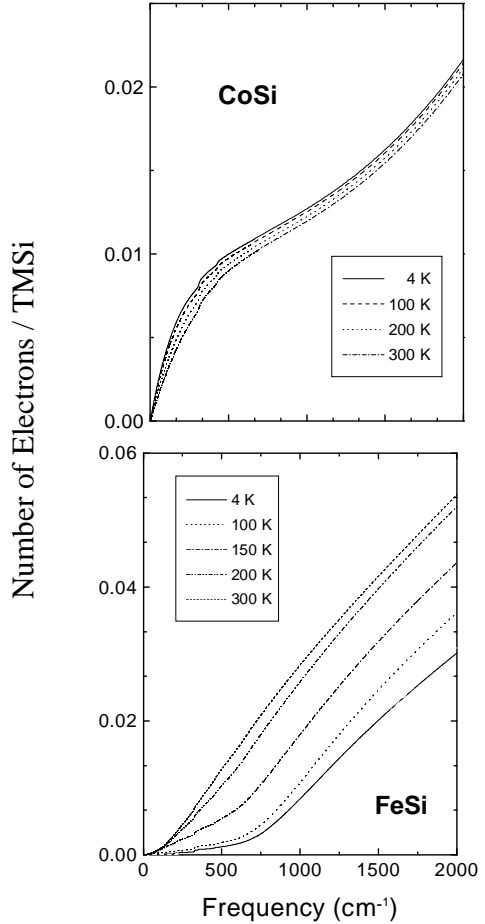}}
\end{center}
\vspace{0cm}
\caption{Number of carriers per TMSi versus frequency ($\omega$) obtained from 
partial integration of the optical conductivity upto $\omega$.}
\label{fig:ne.fesi.cosi}
\end{figure}
\begin{figure}
\vspace{0cm}
\begin{center}
\leavevmode
\vspace{0cm}
\hbox{%
\epsfxsize=75mm
\epsffile{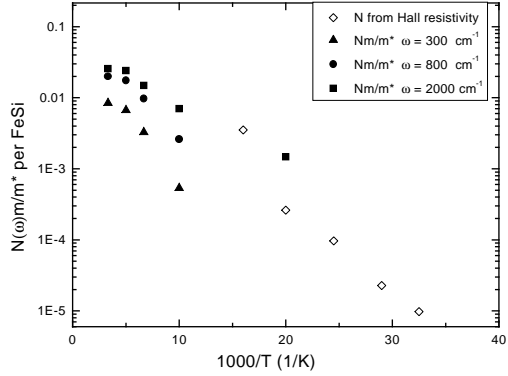}}
\end{center}
\vspace{0cm}
\caption{Number of carriers per unit of FeSi from integrating the optical conductivity 
(closed symbols) and Hall conductivity (open diamonds).}
\label{fig:nel.arr}
\end{figure}

In Fig. \ref{fig:epsig.fesi.cosi} we reproduce 
the real part of the optical conductivity and of
dielectric constants for some selected temperatures of FeSi and CoSi.
In Fig. \ref{fig:ne.fesi.cosi} we display the function 
$Nm_e/m^*(\omega)=2m_e\pi^{-1} q_e^{-2} \int_0^{\omega}\sigma(\omega)$ for
both compounds. The high frequency limit corresponds to
the total number of electrons per FeSi ({\em i.e.} 40 electrons). Here
we neglected the contribution of the nuclei, which is justified due to the 
large ratio of the nuclear mass compared to the electron mass. 
We see, that for CoSi the $Nm_e/m^*(\omega)$-curves for different temperatures
merge above 1000 cm$^{-1}$ indicating that no transfer of spectral weight
from high to low frequencies takes place. 
\\
For FeSi the recovery of spectral
weight can be followed upto approximately 2000 cm$^{-1}$ above which the
experimental errorbars become rather large as a result of uncertainties
in the Kramers-Kronig analysis, mainly induced by the extrapolations
at high frequency\cite{schlesinger,degiorgi}. It seems plausible to associate
the temperature dependence of the section between 1000 and 2000 cm$^{-1}$
with the optical conductivity due to thermally induced charge carriers.
In Fig. \ref{fig:nel.arr} this number is displayed versus temperature along
with the Hall number of Paschen et al \cite{paschen}. We display
$Nm_e/m^*(\omega)$ for 3 different values of the cutoff frequency
($\omega = $ 300, 800 and 2000 cm$^{-1}$) to demonstrate that the
activation energy is independent of frequency. We observe that 
the Hall number and $Nm_e/m^*$ have approximately the same activation
energy of about 200 to 250 K. 
\subsection{Activation energies}
\begin{table}
\begin{displaymath}
\begin{array}{|l|l|l|l|}
\hline
\mbox{Exp. Method} & \mbox{Source} &\mbox{Temp.} & E_a \mbox{(meV)} \\
\hline
\mbox{Tunneling} & \mbox{Ref. \cite{aarts}} & 4 K & 43 \\
\mbox{IR spectr.} & \mbox{Ref. \cite{damascel1}} & 4 K & 38 \\
\mbox{Raman} & \mbox{Ref. \cite{cooper}} & 4 K & 47 \\
\hline
\mbox{K(Si)} & \mbox{Ref. \cite{wertheim}} & < 800 K & 41  \\
\chi(T) & \mbox{Ref. \cite{jaccarino}} & < 800 K & 68  \\
\mbox{Q(Fe)}& \mbox{Ref. \cite{wertheim}} & < 800 K & 46  \\
\mbox{n(Hall)} & \mbox{Ref. \cite{paschen}} &<  100 K & 22  \\
\mbox{n(IR)} & \mbox{This work} & > 100 K & 22  \\
\rho_{DC} & \mbox{This work} & > 100 K & 30  \\
\hline
\end{array}
\end{displaymath}
\caption{Compilation of Activation energies observed for FeSi. For
tunneling, Raman spectroscopy and infrared spectroscopy the indicated
value corresponds to the gap devided by two. K(Si) is the activation
energy of the Knight shift at the Si nucleus, Q(Fe) of the nuclear 
quadrupole  splitting of Fe observed with M\"ossbauer spectroscopy.}
\label{table:ea}
\end{table}
In Table \ref{table:ea} we collect the activation energies and gap 
values determined
from various spectroscopic data and transport properties. For the
tunneling data the gap-value corresponds to the peak-to peak value
in the dI/dV versus V curves. In a single particle band picture this
value should match the optical and Raman gap. Clearly there
is a rather large spread of activation energies. The trend is, that
the higher the temperature range considered, the larger the activation 
energy obtained. This indicates, that the thermal evolution involves a 
distribution of activation energies.
\subsection{Transport properties}
\begin{figure}
\begin{center}
\leavevmode
\hbox{%
\epsfxsize=75mm
\epsffile{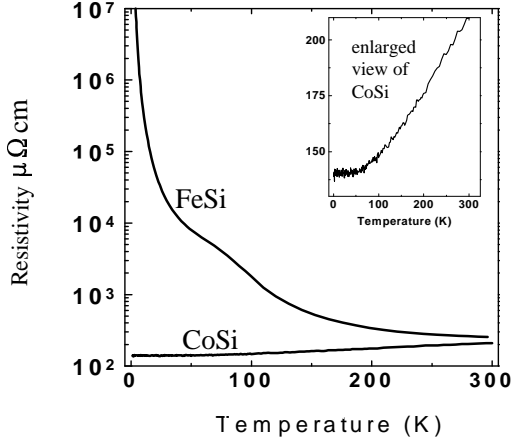}}
\end{center}
\vspace{0cm}
\caption{Resistivity of FeSi and CoSi}
\label{fig:rho.fesi.cosi}
\end{figure}
The behaviour of the electrical 
resistivity of FeSi and CoSi is quite different 
(see Fig. \ref{fig:rho.fesi.cosi}). CoSi has a
high residual resitance and conventional Bloch-Gr\"uneisen
behaviour. FeSi shows insulating behaviour at low temperature. 
The DC resistivity of FeSi $\rho=m^*/(ne^2\tau)$ varies from
2500 to 300 $\mu\Omega cm$ between 100 and 300 K.  When combined
with the carrier densities of Fig. \ref{fig:nel.arr}
we see, that $\tau$ varies from 4.6 to 10 fs between 100 and 300 K. Note 
that the same ratio $n/m^*$ appears in
the Drude expression as in the expression of the optical carrier
density. Hence our result for $\tau$ does not depend on assumptions for 
the effective mass. 
An estimate of the carrier life-time $\tau$ is also obtained from direct
inspection of $\sigma(\omega,T)$. As there are
no thermally excitated carriers at 4K we proceed by subtracting
$\sigma(\omega,4K)$ from the conductivity spectra at higher temperature.
In zeroth approximation the increase in conductivity upon raising
temperature is a Drude-Lorentzian due to the thermally excited 
carriers. At the gap frequency and the optical phonons
the difference spectra contain quirks due to the shifting of the gap, and thermal broadening
of the phonons. Otherwise the difference curves have a halfwidth of 
$ 500 < (2\pi c\tau)^{-1} <1000$ cm$^{-1}$, or $10 fs > \tau > 5 fs$,
in good agreement with the lifetime calculated from the DC resistivity. 
\\
Let us first explore the possibility that the excited carriers can be 
described in the same way as in conventional doped semiconductors. The
carrier density and $k_F$ are related through $n_e=gk_F^3/(6\pi^2)$, where
g is the overall degeneracy factor (valley and spin).
Band theory predicts a 12-fold degeneracy of the electron pockets, and 
an 8-fold degeneracy of the hole pockets\cite{mattheis,tmrice}. 
To simplify matters we will ignore possible differences in
mass and mobility for electrons and holes. Hence the degeneracy factor 
entering the optical carrier density is $g=2\times(8+12)=40$. The observed
density  n$_e$= varies from 0.3 to 1.3 $10^{21}$ cm$^{-3}$ (100 to 300 K), 
implying that $k_F$ varies from $4.9$ to $7.5\cdot 10^6$ cm$^{-1}$, 
$v_F=$ varies from $5.6$ to $8.7 \cdot 10^4$ m/s and $E_F$ varies from 
$9$ to $21$ meV. Hence the electrons and holes are essentially nondegenerate 
throughout this temperature range as expected for thermal excitation across a
semiconductor gap.
Using our estimate of $v_F$ we calculate a mean free path of 
2.6 to 9.0 $\AA$ for temperatures in the range of 100 to 300 K. This
mean free path is sufficiently short to smear the momentum of
the charge carriers on the scale of the size of the electron and hole pockets
($0.02 < k_F l/(2\pi) < 0.1$ ), and even sufficient to smear the
momentum over the entire Brillouin zone ($0.5 < l/a < 2$). This implies
that the conduction of the charge carriers is essentially of the hopping type.
However, this small value of $l$ implies that the fine details of the
bandstructure become irrelevant, and the multi-valleyed character of electron and 
hole pockets is scrambled, thus invalidating the assumption made on input 
that $g=40$.
\\
Let us now compare the absolute values of the carrier density in 
Fig. \ref{fig:nel.arr} obtained from the Hall effect and from the optical
conductivity. 
If we extrapolate the Hall number to the region between 100 and
300 K where we have reliable estimates of the optical carrier density($Nm_e/m^*$),
the Hall number seems to overshoot $Nm_e/m^*$ with a factor 5 to 10.
In a semiconductor model with equal electron
and hole masses and mobilities the optical carrier
density corresponds to $n_{opt}=n_e +n_h$, whereas
$1/n_{Hall}=1/n_e-1/n_h$. There is a compensation between
electron and hole contributions to the Hall constant, leading to an
experimental Hall number which is always larger than the optical
carrier density. 
\\
The combination of electron-like Hall conductivity, and short
mean free path of the carriers, motivates us to propose a different 
scenario from the semiconductor picture which we explored above:
The effect of raising temperature is to excite electrons to a 
wide band above the Fermi energy. The holes which are left behind move
within a narrow band of (almost) localized states. From the perspective
of the effect on doping the latter states behave in the same way as donor
states in a conventional semiconductor, but should be really understood
as a narrow band of quasi-atomic states of the transition metal atoms. 
As we may now ignore the contribution to
the charge response of the hole channel (due to the low mobility) the smaller
value of the optical carrier density $Nm_e/m^*$ compared to the (extrapolated) 
Hall number in the 100-300 K range indicates a moderate effective mass of the 
mobile charge carriers $m^*/m_e\approx 5 - 10$.
An explicit realization of this situation is the pairing of itinerant 
electrons with local moments at low temperature to form a lattice of
Kondo-singlets\cite{schlesinger}. At sufficiently high temperatures these 
singlets will be broken into their constituents, {\em i.e.} itinerant 
electrons and local moments. As no mobile charges are associated with the 
local moments, these moments add a neglegible contribution both to the Hall 
conductivity and the infrared conductivity, so that the dominant contribution
comes from the thermally excitated itinerant electrons. A natural 
consequence of having unpaired local moments at
higher temperatures is spin-flip scattering of the itinerant electrons, 
leading to a reduction of the mean free path. This scattering 
mechanism will be suppressed if the local moments are aligned by the external 
application of a magnetic field (positive magneto-conductivity).
\\
Motivated by these arguments we recalculate the carrier properties,
adopting a degeneracy factor $g=2$ (only spin): $k_F$ now varies 
from $1.3$ to $2.0\cdot 10^7$ cm$^{-1}$, 
$v_F=$ varies from $1.5$ to $2.4 \cdot 10^5$ m/s, 
$E_F$ from 66 to160 meV, $l$ from 7.0 to 25.0 $\AA$, 
and $k_F l/2\pi$ from 0.15 to 0.8.
\subsection{Resonating bonds coupled to phonons}
The steps in the $Nm_e/m^*(\omega)$ curve in Fig. \ref{fig:ne.fesi.cosi}
at frequencies below 1000 cm$^{-1}$ are 
due to optical phonons\cite{damascel2}. By applying 
the optical sum-rule to these vibrational resonances we can 
determine the dynamical (or transverse effective) charge of Fe and Si for 
each temperature. 
For a purely ionic insulator this is also the actual charge of the
ions. For a covalent compound with resonating bonds a finite value of the
transverse effective charge results from a dynamical 
charge redistribution associated with
the ionic motion of an optical phonon. For the TM mono-silicides Pauling
has argued, that resonance occurs between the partial bonds of
types 2 and 3 (1/3 and 2/3 bonds). Lucovsky and White\cite{lucovsky} have shown (for the 
IV-VI narrow gap semiconductors) that this phenomenon leads to a high value 
of the transverse effective charge. Rice, Lipari and Str\"assler\cite{rice} developed
the theory for the infrared spectra for phonons coupled to charge excitations.
\begin{figure}[t]
\begin{center}
\leavevmode
\vspace{0cm}
\hbox{%
\epsfxsize=75mm
\epsffile{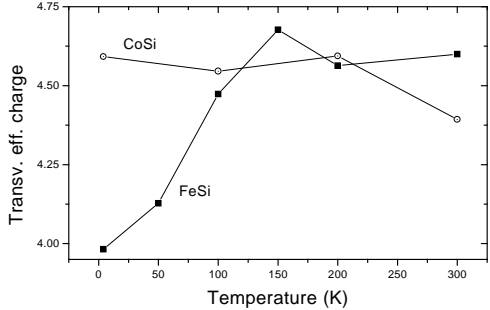}}
\end{center}
\vspace{0cm}
\caption{Transverse effective charge of Fe-Si and Co-Si pairs calculated 
from the oscillator strength of the optical phonons.}
\label{fig:zeff.fesi.cosi}
\end{figure}
In Fig. \ref{fig:zeff.fesi.cosi} we display the transverse effective charge 
for both compounds. A large value of
more than 4 is found for $Z^{eff}$. Based on the electronegativities and
from the observed atomic radii there can be almost no charge transfer 
between the transition metal atom and Si in these compounds. Hence we
attributed the large value of the transverse effective charge to a dynamical 
coupling of the optical phonons to electronic resonances. The full 
analysis was presented in Ref. \cite{damascel1}. 
This requires that at least some of the bonds between Fe and Si are incomplete. The 
enhancement of transverse effective charge is then due to a coupling of the 
ionic displacement to charge flow between the unsaturated bonds.
\section{The electronic structure of TMSi}
\begin{figure}
\begin{center}
\leavevmode
\hbox{%
\epsfxsize=70mm
\epsffile{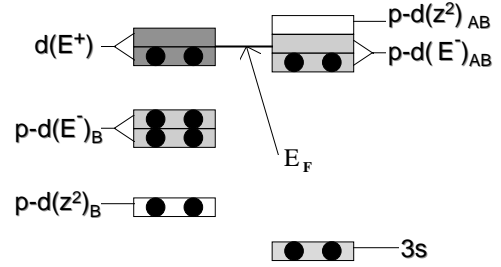}}
\end{center}
\caption{Schematic diagram of the energy levels in TMSi. The situation
with two electrons in the d(E$^+$) level which is displayed here 
corresponds to FeSi.}
\label{fig:energy.levels}
\end{figure}

From ab initio bandstructure calculations for FeSi it is known, that in
these materials the Si-3s states are fully occupied and have a binding energy 
of about 10 eV. The Si 3p states are partially occupied and form together
with the TM-states a complex manifold of bands.
\\
According to a famous rule from the
theory of Mott-Hubbard metal insulator transitions the atomic description of
the TM atom becomes the appropriate Ansatz if the on-site correlation energy 
exceeds the bandwidth. For the TM silicides the bandwidth is of order 6 eV,
and $U^{eff}$ is of the same order of magnitude. The silicides therefore
seem to provide a border case. A band description where the correlation 
effects are taken into account perturbatively has 
been used by several investigators\cite{morya,edwards,doniach,tmrice}. 
In this paper we want to explore a
more intuitive approach based on an atomistic picture.
\\
In their elemental metallic phase the TM atoms have approximately a 
$d^{n-2}s^2$ configuration, with $n$ the number of valence electrons 
per atom. From the d-shell radii displayed in Fig. \ref{fig:radii} we 
can see, that 
in the whole series the d-shell radius\cite{df} falls well below the 
atomic radius, which implies that the d-shells exhibit
quasi atomic characteristics due to shielding by the outer s and p-shells.
Due to the low site symmetry of the TM-atoms mixing of s, p and d-character
occurs in the eigenstates. For the purpose of compactness of notation we will 
label the TM valence bands as 'd'-states. We will assume that these states 
have quasi-atomic characteristics, so that we can use a ligand field 
description to describe the influence of the Si-matrix. 
For the purpose of bookkeeping of the actual d- and sp-count of the atoms,
it is important to take into account the relative contribution of s and p 
character to the 'd'-states. 
\\                          
With this approach we arrive at the following scenario: Each TM atom contributes 
5 d-states and n electrons. In the case of FeSi: n$=$8. Each Si contributes 3 
3p states and 4 electrons. Due 
to the 3-fold rotation symmetry the 5 d-states are grouped in a single 
degenerate $d_0(z^2)$ state, and two groups of doubly degenerate levels 
($E^+$ and $E^-$) corresponding to linear combinations of 
$d_{-2}$,$d_{-1}$,$d_{1}$ and $d_{2}$. Here we adopted a local reference
frame with the $z$-axis oriented along the 3-fold rotation axis which 
pierces the TM atom.
\\
Of the surrounding shell of 7 Si atoms only the 3p orbitals with their lobes 
pointing toward the transition metal atom couple to the TM states. Thus
7 ligand orbitals are engaged in bonds with the TM-atom. We indicate the 
ligand of type 1 as $|L_{0,1}>$. Taking into account the hopping $\tau$ between 
the Si atoms at relative angles $\phi=0$,$\pi/3$ and $2\pi/3$,  
we obtain the ligand-states of all types $j$ ($j=$1,2,3)
\begin{displaymath}
\begin{array}{|l|}
\hline
|L_{0,j}>=|p_0>+|p_{+\eta}>+|p_{-\eta}> \\ 
E_0=\epsilon_p-2\tau\\
|L_{+,j}>=|p_0>+e^{2\pi i/3}|p_{+\eta}>+e^{-2\pi i/3}|p_{-\eta}>\\
E_+= \epsilon_p+\tau\\
|L_{-,j}>=|p_0>+e^{-2\pi i/3}|p_{+\eta}>+e^{2\pi i/3}|p_{-\eta}>\\ 
E_-= \epsilon_p+\tau\\
\hline
\end{array}
\end{displaymath}
The degeneracy between the ligand states $|L_{\pm,j}>$  with j=2 and j=3 is 
lifted due to a finite hopping probality between these groups:\\
\hbox{$<L_{\pm2}|H|L_{\pm,3}> = \tau'$}\\
The dependence on the angular coordinate $\phi$ of the hopping integral between 
the $d_{\pm 2}$ ($d_{\pm 1}$) and the ligand 3p orbital is $e^{\pm 2i\phi}$ 
($e^{\pm i\phi}$). The angle $\eta_j$ identifies the orientation of the two
rings of ligand Si atoms: $\eta_j=0 resp. 24$ degree for type j=2 resp. 3.
The resulting (nonzero) hopping matrix elements between the ligand 
orbitals and the central TM atom are with this choice of local reference frame\\
\hbox{$<d_{0}|H|L_{0,j}> = t_{0,j}$}\\
\hbox{$<d_{\pm 1}|H|L_{\pm,j}> =  t_{1,j}e^{ \mp i\eta_j}$}\\
\hbox{$<d_{\pm 2}|H|L_{\pm,j}> =  t_{2,j}e^{ \pm 2i\eta_j}$} \\
All other matrix elements are zero.
In principle all ligand combinations of the TMSi$_7$ cluster are shared 
between neighbouring TM-sites and form bands. 
Due to the angle of $\simeq$ 70 degrees between the bonds of types
2 and 3, we anticipate that a linear combination of $d_{\pm1}$ and $d_{\pm2}$ exists
which has a strongly reduced hopping matrix element to the cluster
of ligand orbitals, resulting in a (doubly degenerate) non-bonding orbital of a
quasi-atomic character. 
The resulting level scheme
of hybrid TM-ligand character is sketched in Fig. \ref{fig:energy.levels}. 
The single (per TM atom) 
axial bonding orbitals of $d_0 \ - \ L(1)$ character have the highest binding
energy, followed by a band containing the 2 states of bonding 
$d(E^{-}) \ - \ L_{\pm}(2,3)$ character. Around $E_F$ we find a combined partially 
occupied
anti-bonding band, containing  3 states in total. Finally two non-bonding
$d(E^{+})$ states are located at $E_F$. With a finite value for the 
hopping parameter of these states to the ligand orbitals taken into account, the latter 
states form a narrow band. Provided that the width of this band is 
sufficiently small, it behaves as a lattice of quasi-atomic open shells 
obeying Hund's rules. The observation that two quasi-atomic d-states with
two electrons per Fe atom should exist in FeSi was originally made by Pauling,
using a different line of argueing, based on his analyses of the valence bond sums\cite{pauling}.

\begin{figure}
\begin{center}
\leavevmode
\vspace{0cm}
\hbox{%
\vspace{0cm}
\epsfxsize=75mm
\epsffile{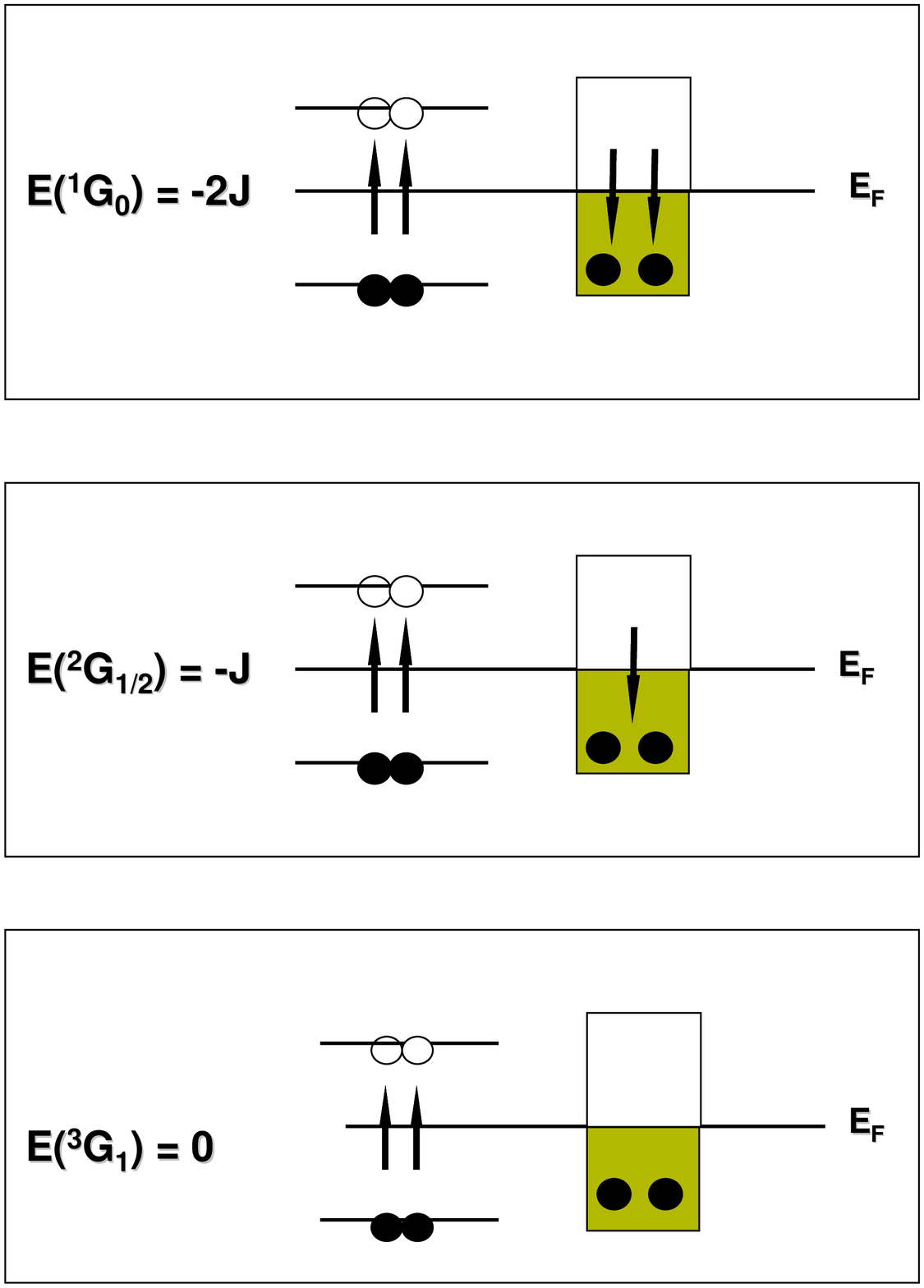}}
\end{center}
\vspace{0cm}
\caption{Schematic diagram of the relevant low energy degrees of freedom of FeSi. The
top panel corresponds to the singlet low temperature phase, the bottom panel
to the high temperature S=1 phase, and the middle panel represents an intermediate situation
with a partial compensation of the local moment.}
\label{fig:kondo.levels}
\end{figure}

 We conjecture, that these subshells are filled 
step by step in the series
CrSi (zero electrons/atom) upto NiSi (four electrons/atom). This leaves
two electrons per Si atom to populate the more itinerant anti-bonding band 
at $E_F$.
This partly filled anti-bonding band corresponds to partial bonds of type
2 and 3, and is responsible for the strong enhancement of the transverse
dynamical charge in CoSi and FeSi\cite{damascel1}. 
\\
This scenario provides a qualitative argument why the same crystal
structure is observed from CrSi upto NiSi: Only if electrons are added to
-or removed from- the itinerant anti-bonding band, chemical bonds are
affected, and the FeSi structure is de-stabilized. Addition
or removal of electrons from this wide band will not take place unless the 
$d(E^{+})$ subshell is either completely full or empty, {\em e.g.} 
for VSi or CuSi. 
It is now easy to understand the high spin state of FeSi at 
elevated temperatures: The electrons in the itinerant band contribute a 
Pauli-paramagnetic susceptibility, which is small.
According to Hund's rules the two electrons in the $d(E^{+})$ subshell are in
a triplet configuration, thus providing the Curie-susceptibility 
with $S=1$ and $g=2$ that has been observed at high temperatures. 
In Fig. \ref{fig:kondo.levels} the relevant elements for the low  energy scale 
are sketched: One local moment with $S=1$ at every site, exchange-coupled to a 
band with two conduction electrons per site. 
\section{Conclusions}
Based on a discussion of infrared and transport data presented in this 
report and of published data, we provide a qualitative model of the electronic
structure of the transition metal monosilicides. The main conclusion is, that
6 electrons per transition metal atom are engaged in chemical bonds
to the Si sublattice. The remaining electrons of the TM atoms (two for FeSi)
occupy a doubly degenerate quasi-atomic d-state, and obey Hunds rules. 
The conduction band supports 2 weakly bound itinerant electrons 
(mainly Si 3p character) which in the case of FeSi compensate the local moment 
at low temperatures by forming a Kondo singlet. This framework provides a 
basis for understanding the observed range of chemical 
stability of these compounds (CrSi to NiSi), and of the physical 
properties of FeSi.
\section{Acknowledgements}
We gratefully acknowledge M. F\"ath for communicating his results to us prior to
publication.
This investigation was supported by the Netherlands Foundation for
Fundamental Research on Matter (FOM) with financial aid from
the Nederlandse Organisatie voor Wetenschappelijk Onderzoek (NWO).


\begin{thebibliography}{999}
\bibitem{aeppli} 
 G. Aeppli, in {\em Strongly Correlated Electron Materials}, 
 edited by K. S. Bedell {\em et al.} 
 (Addison-Wesley, Reading, MA, 1994), p.$\,3$.
\bibitem{mandrus}
 D. Mandrus, J. L. Sarrao, A. Migliori, J. D. Thompson, Z. Fisk, 
 Phys.\ Rev.\ B\ {\bf51}, 4763 (1995).
\bibitem{park}
 C.-H. Park, Z.-X. Shen, A. G. Loeser, D. S. Dessau, 
 D. G. Mandrus, A. Migliori, J. Sarrao, Z. Fisk, 
 Phys.\ Rev.\ B\ {\bf52}, R  $16\,981$ (1995).
\bibitem{xtal} 
 Ru, Os, RhSi: K. G\"oransson, I. Engstr\"om, and B. Nol\"ang,
 J. Alloys and Compounds {\bf 210}, 107 (1995);
 ReSi: R. A. Mc Nees, A. W. Searcy, J. Am. Chem. Soc. {\bf 77}, 4290 (1955).
 Cr, Mn, Fe, Co, NiSi: B. Bor\'en, 
 Ark. Kemi Min. Geol. {\bf 11A}, 1 (1933).
\bibitem{pearson} 
 W. B. Pearson, 
 {\em Crystal Chemistry and Physics of Metals and Alloys},
 Wiley, 1972. 
\bibitem{pauling} 
 L. Pauling and A.M. Soldate, 
 Acta\ Cryst.\  {\bf1}, 212 (1948).
\bibitem{foexradovskiisidorenko}
 F.A.Sidorenko,{\em et al.}, Soviet Phys. J. {\bf 12}, 329 (1969);
 Z.Radovskii {\em et al.}, Soviet Phys. J. {\bf 8}, 98 (1965);
 G. Fo\"ex, J. Phys. Radium {\bf 9}, 37 (1938).
\bibitem{jaccarino} 
 V. Jaccarino, G. K. Wertheim, J. H. Wernick, L. R. Walker, S. Arajs, 
 Phys. Rev. {\bf 160}, 476 (1967).
\bibitem{schlesinger} 
 Z. Schlesinger, 
 Z. Fisk, Hai-Tao Zhang, M. B. Maple, J. F. DiTusa, G. Aeppli, 
 Phys.\ Rev.\ Lett.\ {\bf71}, 1748 (1993).
\bibitem{degiorgi} 
 L. Degiorgi, M. B. Hunt, H. R. Ott, M. Dressel, 
 B. J. Feenstra, G. Gr\"uner, Z. Fisk, P. Canfield, 
 Europhys.\ Lett.\ {\bf28}, 341 (1994).
\bibitem{paschen} 
 M.A. Chernikov, E. Felder, S. Paschen, A.D. Bianchi, H.R. Ott,
 J.L. Sarrao, D. Mandrus, Z. Fisk, 
 Physica B, {\bf 230}, 790 (1997).
\bibitem{aarts} 
 M. F\"ath, J. Aarts, G.J. Nieuwenhuys, A.A. Menovsky and J.A. Mydosh,
 submitted to Phys. Rev. B.
\bibitem{damascel1} 
 A. Damascelli, K. Schulte, D. van der Marel, and A. A. Menovsky,
 Phys. Rev. B {\bf 55}, R4863 (1997).
\bibitem{cooper} 
 P. Nyhus, S. L. Cooper, and Z. Fisk, 
 Phys.\ Rev.\ B\ {\bf51},  $15\,626$ (1995).
\bibitem{wertheim}
 G. K. Wertheim, V. Jaccarino, J. H. Wernick, 
 J. A. Seitchik, H. J. Williams, and R. C. Sherwood,
 Physics Lett. {\bf 18}, 89 (1965). 
\bibitem{mattheis} 
 L. F. Mattheiss and D. R. Hamann, 
 Phys.\ Rev.\ B\ {\bf47},  $13\,114$ (1993).
\bibitem{tmrice}
 V.I.Anisimov, S. Y. Ezhov, I. S. Efimov, I. V. Solovyev, and T. M. Rice,
 Phys. Rev. Lett. {\bf 76}, 1735 (1996).
\bibitem{damascel2} 
 A. Damascelli, K. Schulte, D. van der Marel, 
 J. Mydosh, M. Faeth, J. Aarts, A. A. Menovsky, 
 Physica B {\bf 230}, 787 (1997).
\bibitem{lucovsky}
 G. Lucovsky, and R. M. White,
 Phys.\ Rev.\ B\ {\bf8}, 660 (1973). 
\bibitem{rice}
 M. J. Rice, N.O. Lipari, and S. Str\"assler,   
 Phys.\ Rev.\ Lett.\ {\bf39}, 1359 (1977).
\bibitem{morya} 
 Y. Takahashi, and T. Moriya, 
 J. Phys. Soc. Japan, {\bf 46}, 1451 (1979).
\bibitem{edwards} 
S. N. Evangelou and D. M. Edwards, 
J. Phys. C {\bf 16}, 2121 (1983).
\bibitem{doniach} 
C. Fu, M. P. C. M. Krijn, and S. Doniach, 
 Phys.\ Rev.\ B\ {\bf49}, 2219 (1994).
\bibitem{df}
 D. van der Marel, and G. A. Sawatzky, Phys. Rev. B {\bf 37}, 10674 (1988).
\end{thebibliography}
\end{document}